# Strong Interference HVSR Data Processing and Denoising: HVSR Curve Reconstruction Method based on UPEMD

Bingxuan Song[1], Fuxing Han[1],* Yubei Chen[1], Linjun Wu[1], Mengting Huang[1], Yanjie Pan[1]
[1] the College of Geo-exploration Science and Technology, Jilin University, 130026, China

Abstract

Urban areas pose a challenge for the application of the H/V method due to a high degree of artificial noise. The existing methods fall short in reducing the noise of strong interference data. To solve this issue, a new approach called the HVSR curve reconstruction method is introduced in this paper. The method employs the UPEMD technique to analyze the data component, and the extracted signal is evaluated based on the correlation coefficient between the $IMFs$ and the original micro-motion data, trend extraction of micro-motion data, and secondary extraction. This signal is then utilized to retrieve information about the layers, and the effectiveness of the proposed method is demonstrated.

## 1 Introduction

The horizontal-to-vertical spectral ratio (HVSR) of ambient noise measurements represent a swift and cost-effective non-invasive detection technique. Nakamura[Nakamura(1989)] introduced the estimation of resonance frequency and amplification factor at a specific site. The theory caculates the Fourier amplitude spectrum ratio of horizontal and vertical ground motion components, also known as the H/V spectral ratio. Its robust anti-interference capability has led to its widespread adoption in urban geophysical exploration, without causing surface damage.

Nonetheless, urban exploration faces challenges in dealing with complex human-generated noise sources, including vehicle vibrations and construction

*the College of Geo-exploration Science and Technology,Jilin University

activities. Although it possesses some anti-interference properties, the effectiveness of utilizing raw, unprocessed HVSR data for calculations remains limited. Hence, a signal-to-noise analysis of the original three-component data is essential to derive high-quality spectral ratio curves through thorough processing.

The field of HVSR data signal extraction can be categorized into two technical approaches: identifying transient interference signals in the time domain or detecting outlier fundamental frequencies $(f0)$ in the frequency-domain. The two technical routes will be discussed separately below.

The approach of identifying transient events based on time domain analysis originated from automatic seismic signal picking. Stevenson[Stevenson(1976)] initially proposed the use of the STA/LTA (short time average/long time average) method to detect micro-seismic events in Flathead Lake, Montana. Additionally, Bambang Setiawan et al.[Bambang et al.(2018)Bambang, Jaksa, et al.] . conducted environmental noise measurements in Adelaide, South Australia, selecting the stable segments for analysis. This algorithm is also recommended for time window selection in the implementation guidelines for the H/V spectral ratio method developed by the SESAME (Site Effects Assessment Using Ambient Excitations) team[D(2007)] . Nonetheless, the STA/LTA method exhibits high sensitivity to the selection of the short time window's duration, requiring manual adjustment by the operator. Consequently, the efficacy of the STA/LTA method relies on the operator's experience, making the results susceptible to subjective factors. To address this issue, D'Alessandro et al.[Capizzi(2016)] introduced a hierarchical clustering algorithm for automating time window selection. This algorithm operates by extracting self-consistent clusters within the HVSR curves, eliminating subjective factors related to threshold and time window length selection within the STA/LTA method, and enhancing the objectivity of data analysis.

In recent years, the lognormal distribution of the fundamental frequency $(f0)$ in the HVSR method, along with the examination of viability, has led to a gradual shift from the time domain approach to the frequency domain extraction method for refining HVSR curve. Brady R. Cox, Tianjian Cheng et al. [Manuel(2020)] comprehensively explored the connection between the fundamental frequency and the HVSR curve's viability. They compared the fundamental frequency with the mean and standard deviation of the fundamental frequency, after fully discussing the usable relationship between fundamental frequency and HVSR curve .Their work introduces a frequency-domain window-rejection algorithm and highlights a shortcoming in the time domain method: the absence of an inherent link between transient disturbances and fundamental frequency outliers in the curves. It also underscores that fundamental frequency outliers may manifest in the spectral ratio signal of a long-term stable time window. Examples of dealing with multi-peak HVSR curves are not common. G. Dal Moro and G.F. Panza[Dal Moro(2022)] conducted a statistical assessment of three cases involving multi-modal HVSR spectral lines. They applied the concept of multi-window fundamental frequency mean and standard deviation to distinguish fundamental frequency outliers within time windows. Their work also highlights the absence of a clear connection between HVSR outliers and channel amplitude.

Furthermore, Lanbo Liu et al.[Liu et al.(2015)Liu, Mehl, et al.] suggested utilizing the Hilbert-Huang transform (HHT) to analyze micro-motion data, leveraging its applicability to non-stationary signal processing. This method decomposes the original signal into IMFs (Intrinsic Mode Functions) through EMD (Empirical Mode Decomposition) and denoises the signal by eliminating the highest-frequency Intrinsic Mode Function (IMF-1) .Then caculate the Hilbert-Huang transform.

In a more recent development, Han Fuxing and Song Bingxuan et al.[Han et al.(2023)Han, Song, et al.] introduced a multiple weighted HVSR denoising method based on XGBoost. This approach combines two key characteristics of HVSR curves: signal stability and peak prominence. Notably, it can autonomously extract signals after model training, diminishing the impact of subjective factors in HVSR curve extraction.

Empirical evidence demonstrates that the previously mentioned frequency domain algorithms, the HHT method, and the multiple weighted HVSR denoising method based on XGBoost are highly effective for data collected in less disturbed field and campus environments, supported by strong theoretical foundations. However, in urban settings, particularly in developing cities characterized by frequent construction and severe industrial noise pollution, the data should be categorized as 'strong interference data' . Data collected during quiet periods, such as at night, warrants special consideration (T the definition of strong interference data will be discussed specifically in the following Need for strong interference data definition and UPEMD application). All of the aforementioned methods prove ineffective in extracting micro-motion data under conditions of strong interference. Brady R. Cox, Tianjian Cheng et al. [Manuel(2020)], and G. Dal Moro, G.F. Panza[Dal Moro(2022)] have emphasized the importance of describing HVSR in the form of statistical reports. G. Dal Moro and G.F. Panza [Dal Moro(2022)] contend that HVSR curves should be treated as data to be correctly interpreted and evaluated, rather than serving as a source of objective values. The SESAME (Site Effects Assessment Using Ambient Excitations) team [D(2007)] also outlines certain interpretation criteria.In practice, Sayed S.R. Moustafa et al.[Khan et al.(2021)] uses a clustering algorithm to perform statistical analysis of the fundamental frequency. On top of that, an approach are expected to be proposed to facilitate the development of stratigraphic inversion by providing dependable and universally applicable data processing techniques.

In light of the issues mentioned above and the evolving direction of the HVSR method, this paper initially addresses the imperative need for defining strong interference data and introduces the UPEMD algorithm in the context of HVSR data processing. Subsequently, the paper presents a method called HVSR curve reconstruction method to achieve the extraction of signals from strong interference data. Lastly, the study applies this method to micro-motion data collected in the construction environment of Shenzhen's Futian district, demonstrating its effectiveness through the inversion of formation data.

# 2 Need for strong interference data definition and UP-EMD application

## 1. Strong interference data definition discussion

The data whose noise pollution amplitude is higher than the micro-motion signal is defined as the strong interference micro-motion data, as shown in Fig.1. In addition to micromotion data dealing with common large-amplitude transient events, strong disturbance data also face large-amplitude long-period events. Due to the complex background noise and large pollution noise energy of such data collection environment, there are also more high-frequency noise pollution. The micromotion data with strong interference may have the phenomenon that the fundamental frequency is hidden and cannot be extracted, which makes the frequency domain algorithm that depends on the fundamental frequency identification unavailable. At the same time, the pattern of strong interference data is completely different from that of ordinary micro-motion data, and the zeros of different time Windows may jump, which makes the method of identifying transient interference completely invalid. (although this method has also been shown to be not entirely correct for normal data). Although the EMD method can separate the strong interference amplitude of the data into the IMF with low frequency, the traditional EMD method takes the micromotion signal after removing the intrinsic mode function with the highest frequency (IMF-1) as the effective signal for extraction. In the strong interference data, the effective signal is likely to be decomposed into the IMF-1, and this method is completely invalid. Even if the valid signal exists in the subsequent IMF, the method how to define what is interference and what is data is also not given. In addition, the EMD method has serious mode splitting effect. In the face of complex interference, how to suppress the mode splitting of effective data is also worth considering (the Need for UPEMD application will be discussed in the following).

Therefore, it is difficult to use any conventional method for effective signal extraction of strongly disturbed micromotion data, which is also the reason for defining strongly disturbed data separately. Its acquisition environment, raw data composition and subsequent processing methods are different from any existing ordinary micro-motion data with numerical examples.

## 2. Need for UPEMD application

The EMD method, originally introduced by Huang et al.[Huang et al.(1998)Huang, Shen, et al.], can decompose the zero-shift phenomenon in the time-window caused by long-term high-amplitude noise into several low-frequency IMFs, so the method has a high application prospect in the analysis of high-noise fretting data. However, EMD is known to suffer from a phenomenon referred to as 'mode mixing'.In the context of HVSR data processing, this phenomenon manifests as the existence of meaningful micro-signals in multiple IMFs simultaneously. In complex data acquisition environments with high interference, the number of IMFs containing valid signals across the three data components may not always be same, as elaborated upon in the special cases discussed below. Consequently, concentrating valid data within one or a few IMFs, can effectively suppress the mode mixing phenomenon. In the existing research, the main means to uppress the mode mixing effect is noise-assisted method.

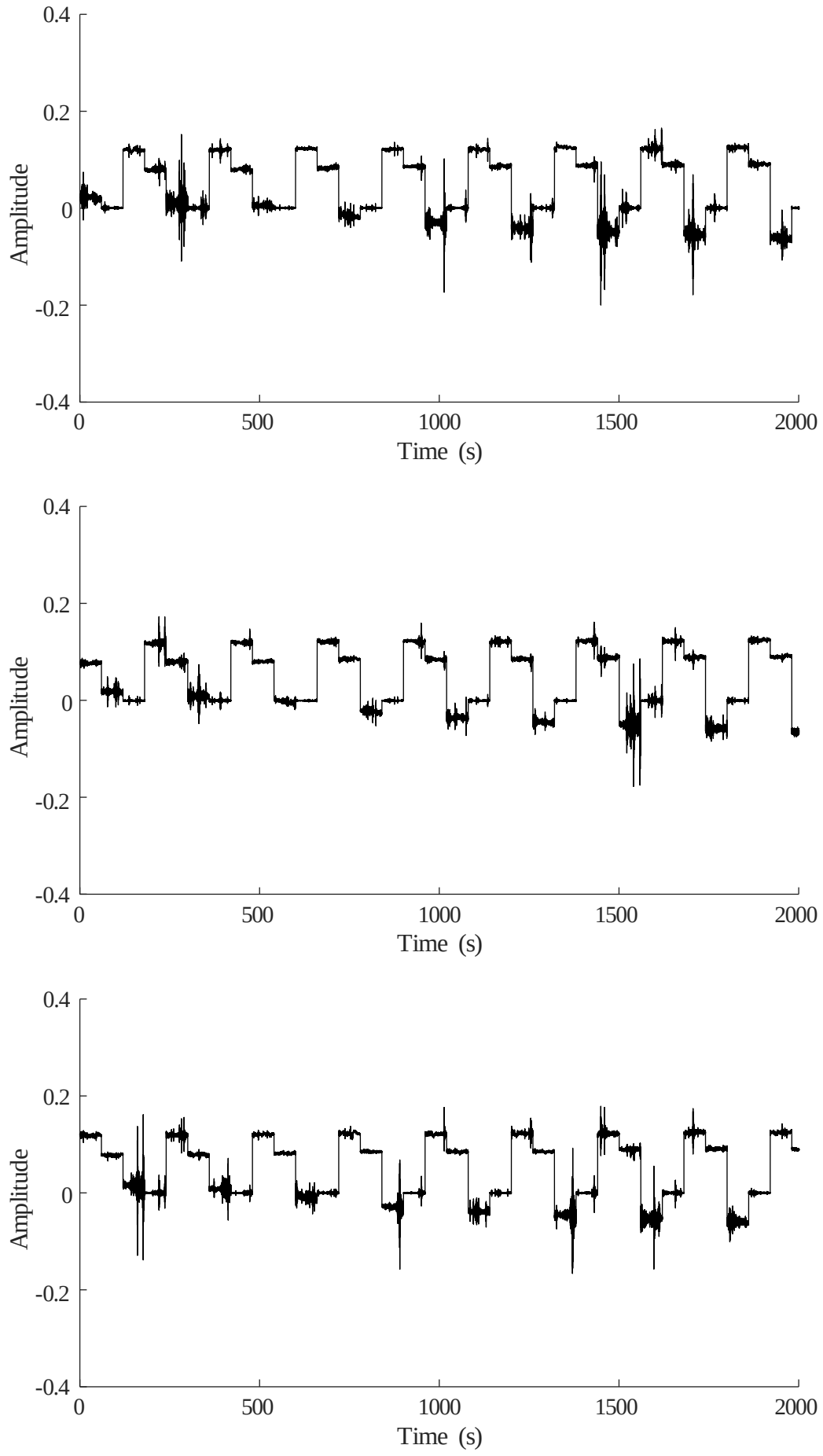

Figure 1: Strong Interference Data

Traditional noise-assisted EMD methods, such as EEMD method, use white noise to suppress the modal splitting phenomenon generated by intermittent high-frequency components, but cannot eliminate the component of auxiliary noise in IMF. If this method is applied to the analysis of frets signal or even the calculation of spectral ratio, the contribution of the auxiliary noise component and frets signal to the spectral ratio value cannot be measured. It is difficult to ensure the physical basis of the spectral ratio method. Among the noise-assisted methods, UPEMD method, as proposed by Wang et al.[YH et al.(2018)YH, K, & MT], fully considers the influence of auxiliary signals, and eliminates the energy contribution of auxiliary noise better while suppressing the mode mixing effect. Therefore, UPEMD method will be adopted in this method to decompose the fretting data.

# 3 Theroy

## 1. Uniform phase EMD

As mentioned above, we adopt UPEMD to process the micro-motion data.

To commence, initial conditions must be established:

$$f_w = 1/T_w \tag{1}$$

$$T_w(m) = 2^m, m = 1 : n_{imf} \tag{2}$$

$$n_{imf} = log_2(n) \tag{3}$$

$$r_0(t) = sgy - data(t) \tag{4}$$

$$\epsilon_m = \epsilon_0 \cdot std(r_m(t)) \tag{5}$$

In the above equation, $n$ is the length of the HVSR data to be decomposed, $sgy - data(t)$ is the HVSR data to be decomposed, and the number of IMFs after UPEMD should be roughly equal to $n_{imf}$. $T_w(m)$ is the period of the masking signal added during the decomposition of the $m^{th} IMF$ , and $f_w$ is its corresponding frequency. $\epsilon_m$ is the masking signal amplitude, and the coefficient $\epsilon_m$ used in the initial value definition depends on experience selection, usually $0.2 \sim 0.4$ is selected.

Caculate masking signal $\omega(t)$:

$$\omega_{k,m} = \epsilon_m \cdot \cos(2\pi \cdot f_w t + \theta_k) \tag{6}$$

$$\theta_k = \frac{2\pi \cdot (k - 1)}{n_p}, k = 1 : n_p \tag{7}$$

In the above equation, $n_p$ is the number of terms, which depends on the empirical selection. The more the number of terms is, the stronger the suppression of mode mixing effect will be, and more calculation time will be consumed. Let $\theta_k$ be the phase of the $k^{th}$ masking signal.

Then we can caculate the Two-Level UPEMD (2L-UPEMD). 2L-UPEMD is the basic cyclic unit of UPEMD. For simplicity, 2L-UPEMD for the $m^{th}$ cycle will be abbreviated as operator $TU_m()$ when used in the following.

**Algorithm 1** Two-Level UPEMD (Take the $m^{th}$ loop as an example)

1: Based on (6) and (7), compute the $m^{th}$ signal to be decomposed:
$y_{k,m}(t) = r_{m-1}(t) + \omega_{k,m}(t)$
2: Preform the EMD to obtain the $IMFs$:
$c_{k,m}(t) = E(y_{k,m}(t))$
The operator $E()$ means decomposing only one $IMF$ from $y_{k,m}$. Then give the $IMF$ magnitude to $c_{k,m}$ .
3: Repeat Step 1 to 2 for $k = 1$ to $n_p$.
4: The $IMF_m$ is obtained by averaging as:
$IMF_m = \frac{1}{n_p} \cdot [\sum_{k=1}^{n_p} (c_{k,m}(t) - \omega_{k,m}(t))]$

As can be seen from the above process, the operator $TU_m()$ will automatically add masking sgnial to the input signal and output an $IMF$ with the masking noise contribution removed.

In the following, 2L-UPEMD is extended to UPEMD:

**Algorithm 2** UPEMD (Also called Multi-Level UPEMD)

1: Assign $n_p$ and $\epsilon_0$
Based on (3) and (4), Set:
$n_{imf} = log_2(n)$
$r_0(t) = sgy - data(t)$
2: Preform the2L-EMD to obtain the IMF:
$IMF_m = TU_m(r_{m-1}(t))$
3: Calculate residue $r_m(t) = r_{m-1}(t) - IMF_m$
4: Repeat Step 2 to 3 for $m = 1$ to $n_{imf}$ to obtain all $IMFs$

For the convenience of expression, UPEMD is abbreviated as operator $U()$, namely $IMFs = U(sgy - data(t))$.

## 3.2 HVSR curve reconstruction method

Decomposing the original micro-motion data $sgy - data(t)$ yields:

$$IMFs = U(sgy - data(t)) \tag{8}$$

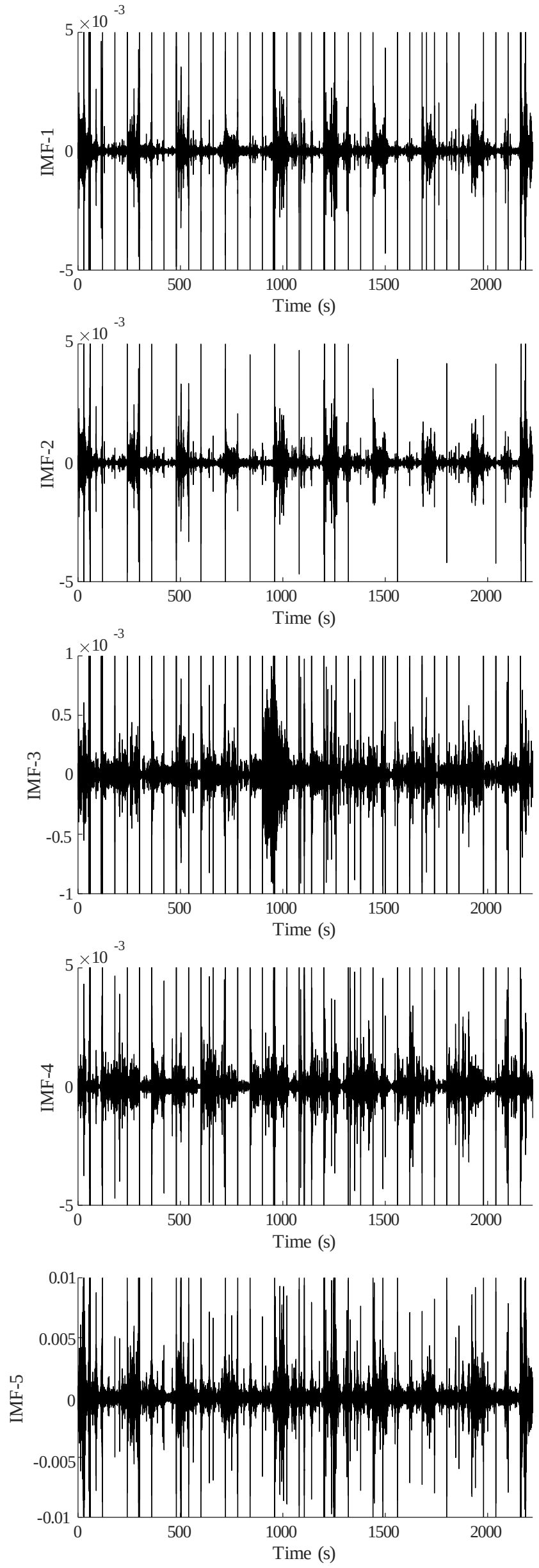

Figure 2: UPEMD Sample

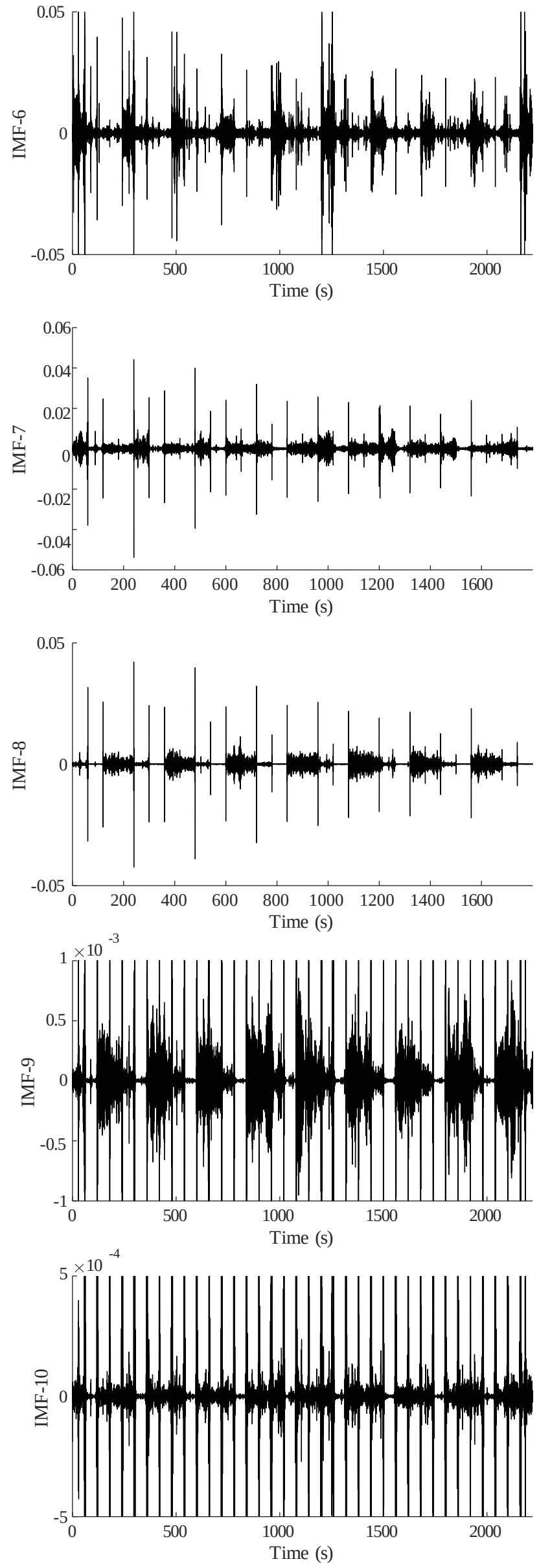

Figure 3: UPEMD Sample

As is shown in Fig.2 and Fig.3,The raw data is decomposed into multiple IMFs. Then the physical meaning of $IMFs$ components is analyzed, and the correlation coefficient between each $IMF$ and the original data is obtained:

$$corr(IMF(i), sgy - data) = \frac{cov(IMF(i), sgy - data)}{\sigma_{IMF(i)} \cdot \sigma_{sgy-data}} \tag{9}$$

In the above equation, $cov(IMF(i), sgy - data)$ represents the covariance between each $IMF$ and the original data, $\sigma_{IMF(i)}$ and $\sigma_{sgy-data}$ represent the standard deviation of each $IMF$ and the standard deviation of the original data, respectively.

The $IMF$ with correlation coefficient greater than 0.01 is selected to obtain $IMF^{1th}$. Because the energy of the large amplitude and long period noise of the strong interference micro-motion data is much higher than that of the effective signal, the correlation coefficient of the $IMF$ dominated by the noise data is greater than 0.01, and experience shows that the correlation coefficient is usually larger than the correlation coefficient of the IMF dominated by the effective signal. Due to the characteristics of transient interference or small amplitude interference with low energy and strong randomness, it is usually included in the highest frequency $IMF(1)$, and the correlation coefficient with the original signal is small. Therefore, there is usually no $IMF(1)$ in $IMF^{1th}$. (The case that $IMF(1)$ exists in $IMF^{1th}$ will be explained in the discussion of special cases) that is, $IMF^{1th}$ is mainly composed of effective micro-motion signal and long-period large-amplitude noise.

In strong interference data, the long period and large amplitude noise pollution amplitude is too high, which can be regarded as a complex trend of the data. Therefore, Gaussian smoothing filter can be used to extract the trend. That is:

$$smoothed - data(t) = \frac{\sum_{t-l_0/2}^{t+l_0/2} f(t_i) \cdot sgy - data(t_i)}{\sum_{t-l_0/2}^{t+l_0/2} (f(t_i)} \tag{10}$$

$$f(t) = \frac{1}{\sigma\sqrt{2\pi}} exp(-\frac{(t_i - t)^2}{2\sigma^2}), t_i = t - \frac{l_0}{2} : t + \frac{l_0}{2} \tag{11}$$

In the above equation, Equation (10) is the Gaussian filter implementation method, and Equation (11) is the window function of the moving smoothing window.

Where, $l_0$ is the width of the smoothing window and $\sigma$ is the standard deviation of the original data within the smoothing window. The essence of equation (11) is the probability density function of normal distribution (normpdf). In fact, most linear filters with poor detail preservation can be achieved because the micromotion data is regarded as noise in this processing step and the polluted part is extracted with emphasis. When dealing with strong interference micro-motion data, due to the normal distributed curve shape of the Gaussian filter, the effective micro-motion signal is filtered while the jump part of the data trend is well preserved, and the sensitivity to the width of the time window is low, and the parameter adjustment is relatively easy.

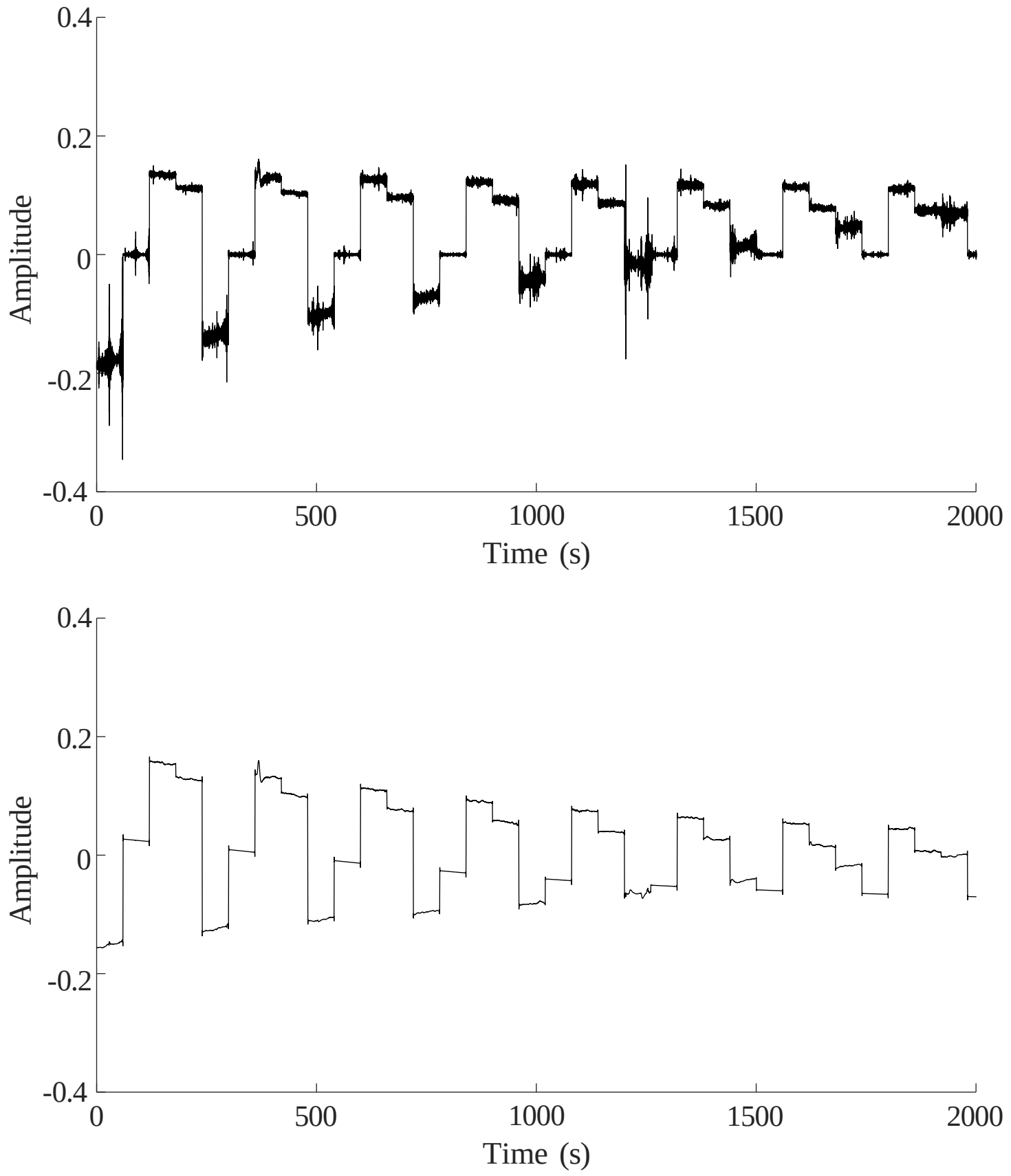

Figure 4: Orginal Data and Reconstructed Noisy Signal

For ease of presentation, the Gaussian smoothing filter will be abbreviated as $G()$ in the following.

Calculate the correlation coefficient between each $IMF$ and the first smoothed data $smoothed - data_1(t)$ :

$$\begin{aligned} & corr(IMF^{1th}(i), smoothed - data_1) \\ & = \frac{cov(IMF^{1th}(i), smoothed - data_1(t))}{\sigma_{IMF^{1th}(i)} \cdot \sigma_{sgy-data}} \end{aligned} \tag{12}$$

The $IMF^{2th}$ is obtained by deleting the $IMFs$ whose correlation coefficient with the first smoothed data is greater than 0.01 from the $IMF^{1th}$. The original micromotion data is mainly composed of three parts: high frequency and low amplitude interference, low frequency and high amplitude interference and effective signal. Low frequency and high amplitude interference contributes the main energy of The $IMF$ whose correlation coefficient with the first smoothed data is greater than 0.01. Therefore, the correlation coefficient values with the original data mainly come from its energy contribution to the low frequency and high amplitude noise. After deletion, the large amplitude low-frequency noise pollution of data will be greatly reduced.Fig.4 shows the orginal data and its noise reconstructed by UPEMD and correlation coefficient solution.

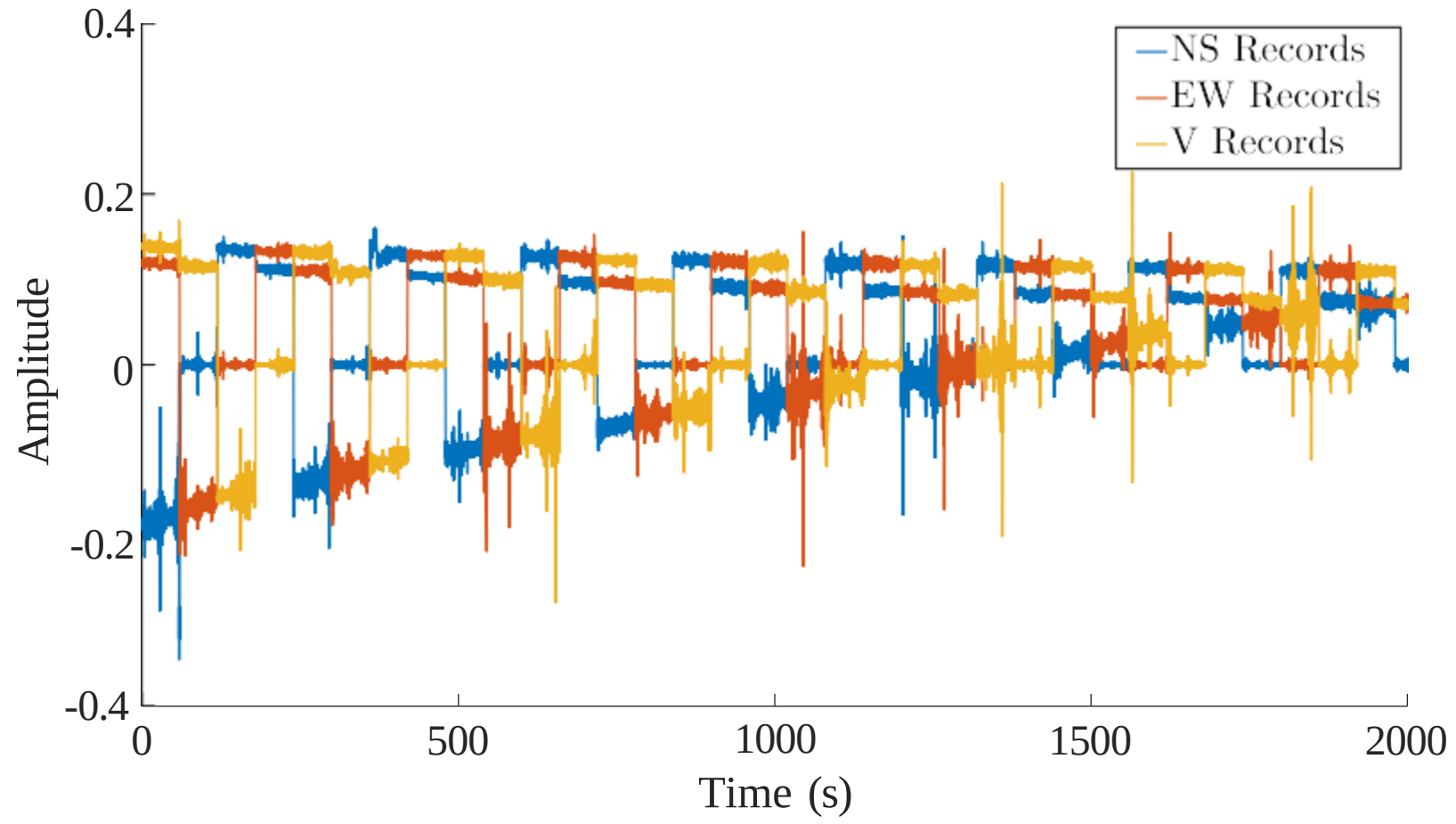

Figure 5: Orginal data

Due to the complexity of the low frequency and high amplitude components, the above processing method can only remove the noise which is far higher than the effective signal order of magnitude, and the remaining interference is difficult to remove. Therefore, the above process is recycled, as follows:

$$smoothed - data_2 = G\left(\sum IMF^{2th}(i)\right) \tag{13}$$

$$\begin{aligned} & corr(IMF^{2th}(i), smoothed - data_2) \\ & = \frac{cov(IMF^{2th}(i), smoothed - data_2)}{\sigma^{2th}_{IMF}(i) \cdot \sigma_{sgy-data}} \end{aligned} \tag{14}$$

The correlation coefficient of $IMF$ greater than 0.01 was selected to obtain the $IMF^{3th}$. $IMF^{3th}$ has eliminated almost all interference. In general, the number of $IMFs$ in the $IMF^{3th}$ of the three-component data is equal and the position in the original $IMF$ is the same. At this point, the micromotion signal can be reconstructed by summing the $IMF^{3th}$ of each component.

## 3. Discussion of special cases

1. Case 1

$$corr(IMF(1), sgy - data) > 0.01$$

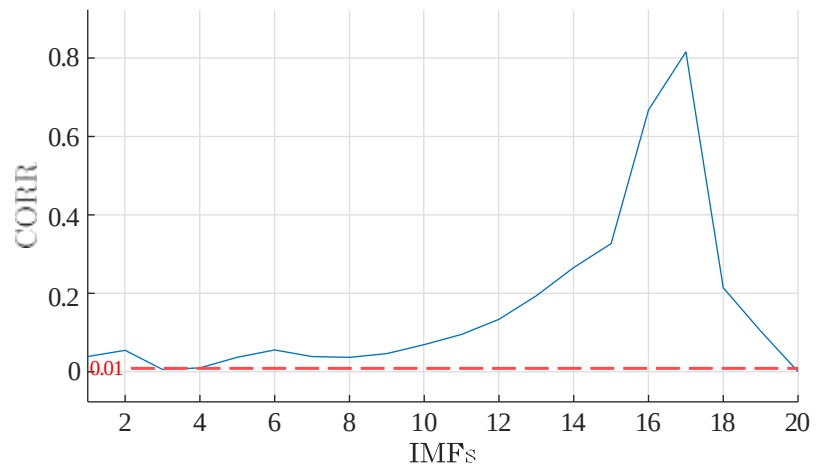

[a]

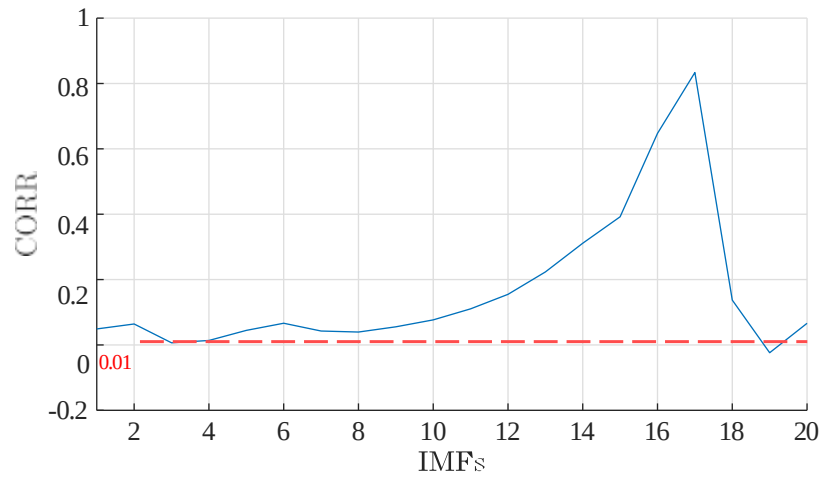

[b]

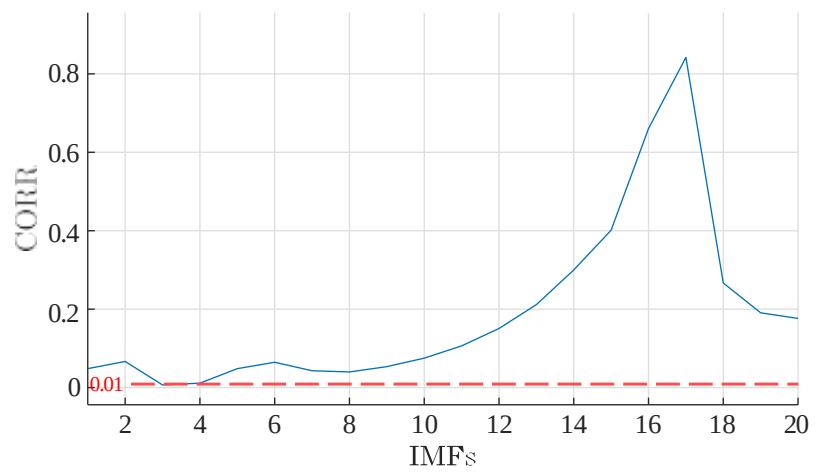

[c]

Figure 6: Correlation Coefficient with the Original Data
Graphs a, b, and c show the correlation coefficients in the NS, EW, and V directions, respectively

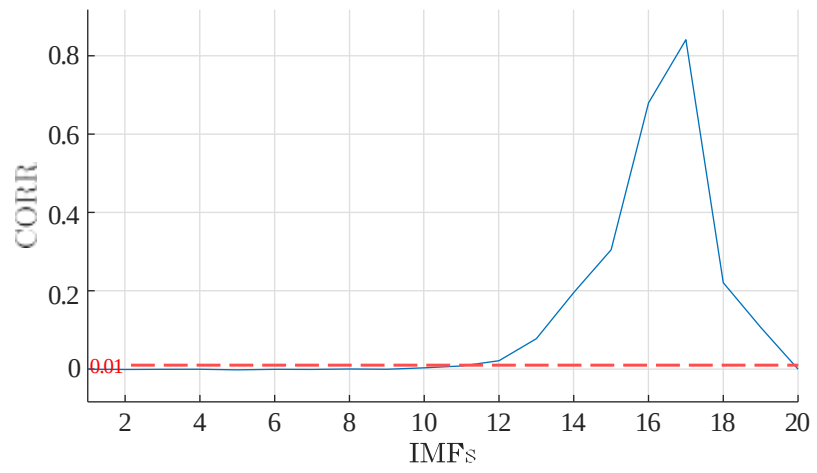

[a]

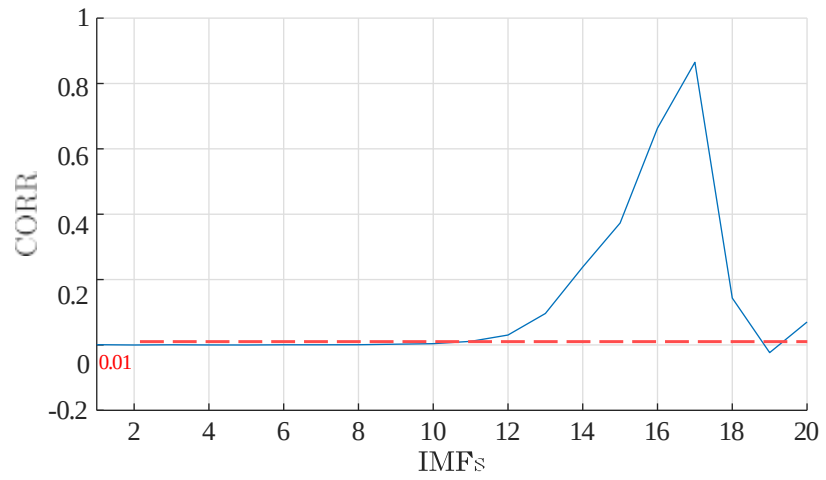

[b]

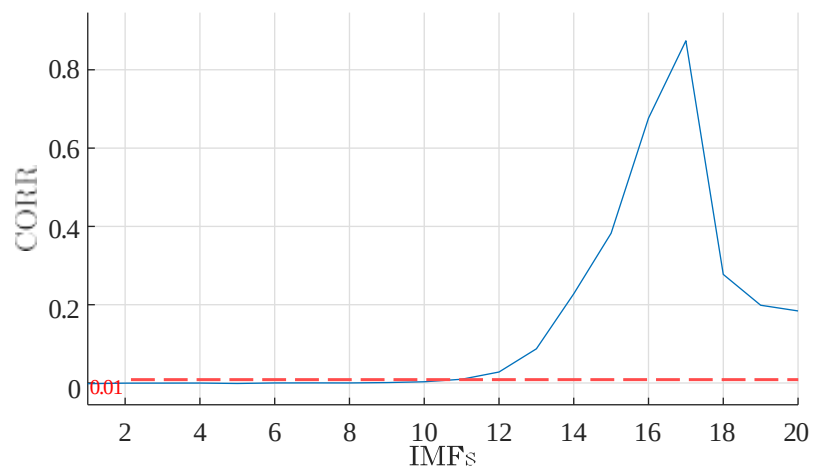

[c]

Figure 7: Correlation Coefficient with the first Smoothing Result
Graphs a, b, and c show the correlation coefficients in the NS, EW, and V directions, respectively

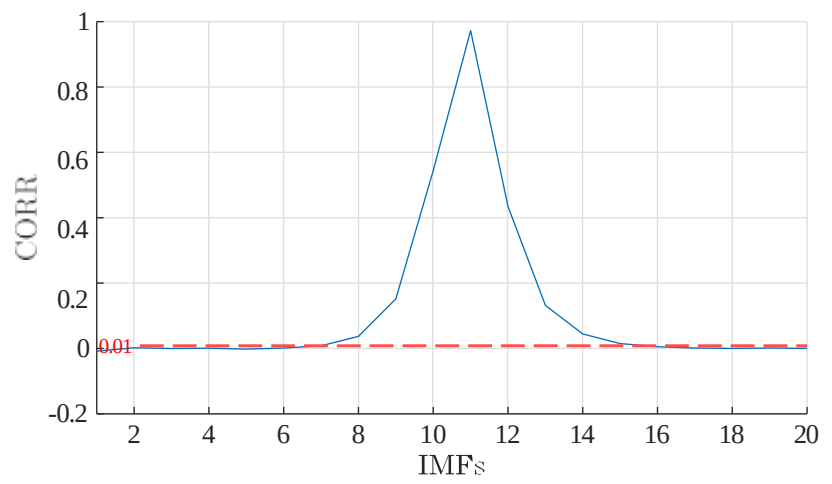

[a]

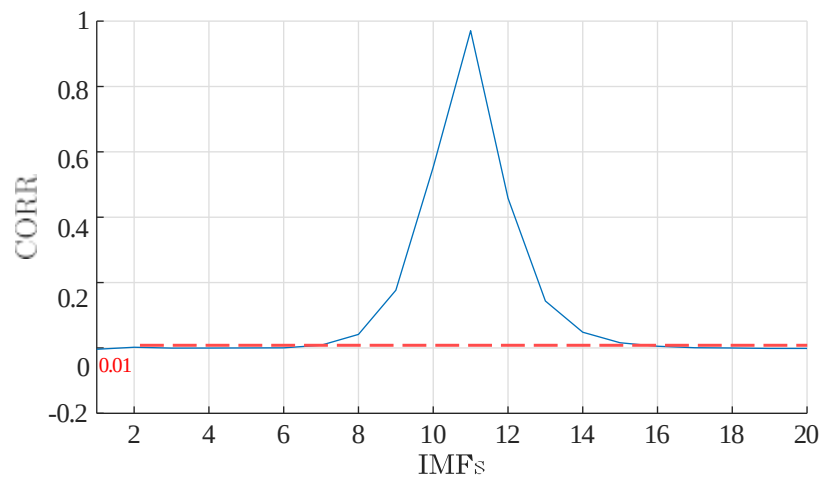

[b]

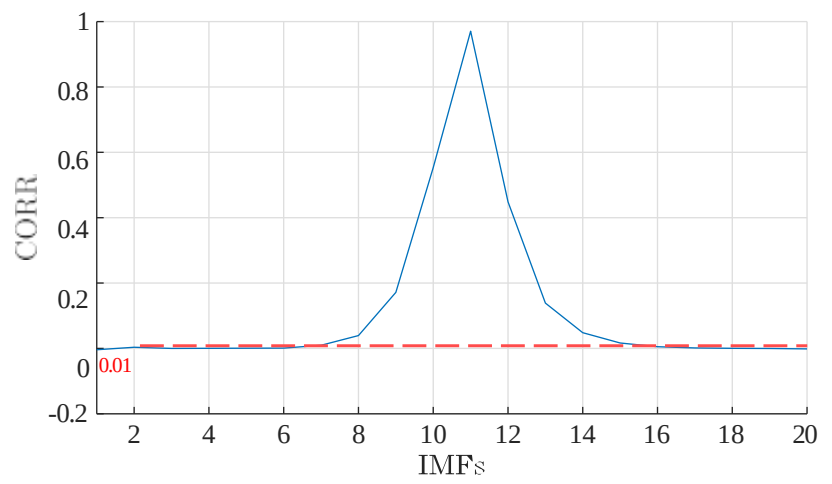

[c]

Figure 8: Correlation Coefficient with the Secondery Smoothing Result
Graphs a, b, and c show the correlation coefficients in the NS, EW, and V directions, respectively. They look very similar, but the values are not the same. This similarity is also a reaction of the validity of the method.

**Algorithm 3** UPEMD (Also called Multi-Level UPEMD)

1: Assign $n_p$ , $\epsilon_0$ and $n_{imf}$.
2: Preform the UPEMD to obtain all the $IMFs$.
3: Calculate the correlation coefficients with the orginal data.
4: Preform the Gaussian smooth to obtain the first smoothed data.
5: Calculate the correlation coefficients with the first smoothed data.
6: Filter the data to obtain the second signal based on the principle above.
7: Preform the Gaussian smooth to obtain the second smoothed data.
8: Calculate the correlation coefficients with the second smoothed data.
9: Determine whether it is a special case and decide to output or change the value of $\epsilon_0$ to redo it.

$$corr(IMF(1), sgy-data) < corr(IMF(2), sgy-data)$$

It is suggested to re-adjust the parameter $\epsilon_m$. If this is still the case, as UPEMD cannot completely suppress the mode splitting effect, the correlation coefficient between $IMF(1)$ and the original signal is greater than 0.01 because the effective signal in $IMF(2)$ or $(3)$ near $IMF(1)$ is leakage into $IMF(1)$. However, $IMF(2)$ or $(3)$ and other subsequent $IMF$ dominate the curve shape, so IMF(1) can be omitted.

### 3.3.2 Case 2

$$corr(IMF(1), sgy-data) > 0.01$$

$$corr(IMF(1), sgy-data) > corr(IMF(2), sgy-data)$$

This situation needs to be combined with the specific analysis of micromotion data collection environment.

The micromotion data acquisition environment has less interference, probably because the data itself is less disturbed by high-frequency low amplitude noise. Practice has proved that $IMF(1)$ dominates the shape of the spectral ratio curve, and $IMF(1)$ can be regarded as effective data for calculation.

When the micromotion data itself belongs to the strong interference data, this situation needs special attention. The strong interference data itself represents the acquisition environment with high noise energy and complex composition, and the data must be affected by high-frequency low-amplitude transient interference. However, compared with the strong interference, the high-frequency low-amplitude transient interference is a minor interference and cannot be removed from the micromotion data. It is suggested to adjust $\epsilon_m$ to calculate again. If this situation still occurs, $IMF(1)$ is the combination of effective signal and transient interference, and $IMF(2)$ means that the IMF with subsequent correlation coefficient greater than 0.01 is composed of effective signal leakage in $IMF(1)$. Therefore, $IMF(1)$ must be included in the valid data for calculation, otherwise it may lead to layer missing during inversion. The above process cannot guarantee the success of data extraction, so it is not recommended to use this data.

3. Case 3

The three-component micro-motion data $IMF^{3th}$ have different numbers or do not correspond to their positions in the original $IMF$

Because the interference of three-component broadband microseismograph in three directions cannot be guaranteed to be completely consistent, the decomposition effect and the mode mixing suppression effect of UPEMD on three-component data is not exactly the same. The non-corresponding position in the original $IMF$ may be due to the pollution of a certain component by the unique interference data of the component. The non-corresponding quantity may be due to the different noise pollution degree of the original data and the different mode mixing suppression effect, which leads to the different number of $IMFs$ dominated by the effective data. In case that an $IMF$ is deleted directly, if the micro-motion data are relatively complete, it may have no effect on the inversion; if the micro-motion data are relatively incomplete, it may lead to the loss of layer information during inversion.

If the effective $IMF^{3th}$ cannot be retrieved from the micro-motion data through the above process, the filtered IMF in $IMF^{2th}$ by the second correlation coefficient solution can be checked separately. In a few cases, because the effect of the first filtering is too good, there is almost no large amplitude low-frequency noise in $IMF^{2th}$.In the process of second filtering and correlation coefficient solving, it is possible that the effective signals are filtered out.

If the above supplementary analysis still fails to retrieve the valid $IMF^{3th}$, it proves that this group of data is too polluted by noise, and it is recommended to be discarded.

4. Additional Notes

Because the decomposition effect of UPEMD method is greatly affected by the transcendental parameter $\epsilon_m$, there is a possibility that high-frequency interference signals can not be eliminated in the $IMF$, and the effective signal can not dominate HVSR curve shape. Therefore, it is recommended that experienced HVSR processing personnel use this method for processing.

# 4 Numerical example

In order to further test the signal extraction effect of this method, the data in the construction environment of Shenzhen Futian District are used for testing.Futian District, under the jurisdiction of Shenzhen City, Guangdong Province, is the central urban area of Shenzhen City, the resident of Municipal Party Committee and Municipal government, and the center of administration, finance, culture, business and international exchanges of the whole city. There are a lot of artificial earth filling in this area, so the early warning of urban cave-ins is very important. HVSR method is used here as a fast and economical non-invasive detection method. But the above human environment determines that the human noise interference in this area will be very intense. Therefore, the data in this area is used as the data source of strong interference. On the one hand, the data in this area conforms to the definition of strong interference data, and on the other hand, this method will be beneficial to the early warning of ground collapse in this area. The results of the UPEMD method for the NS records data are shown in Fig.12(Due to space problems, the data decomposition results of EW and V directions are not shown in this paper).

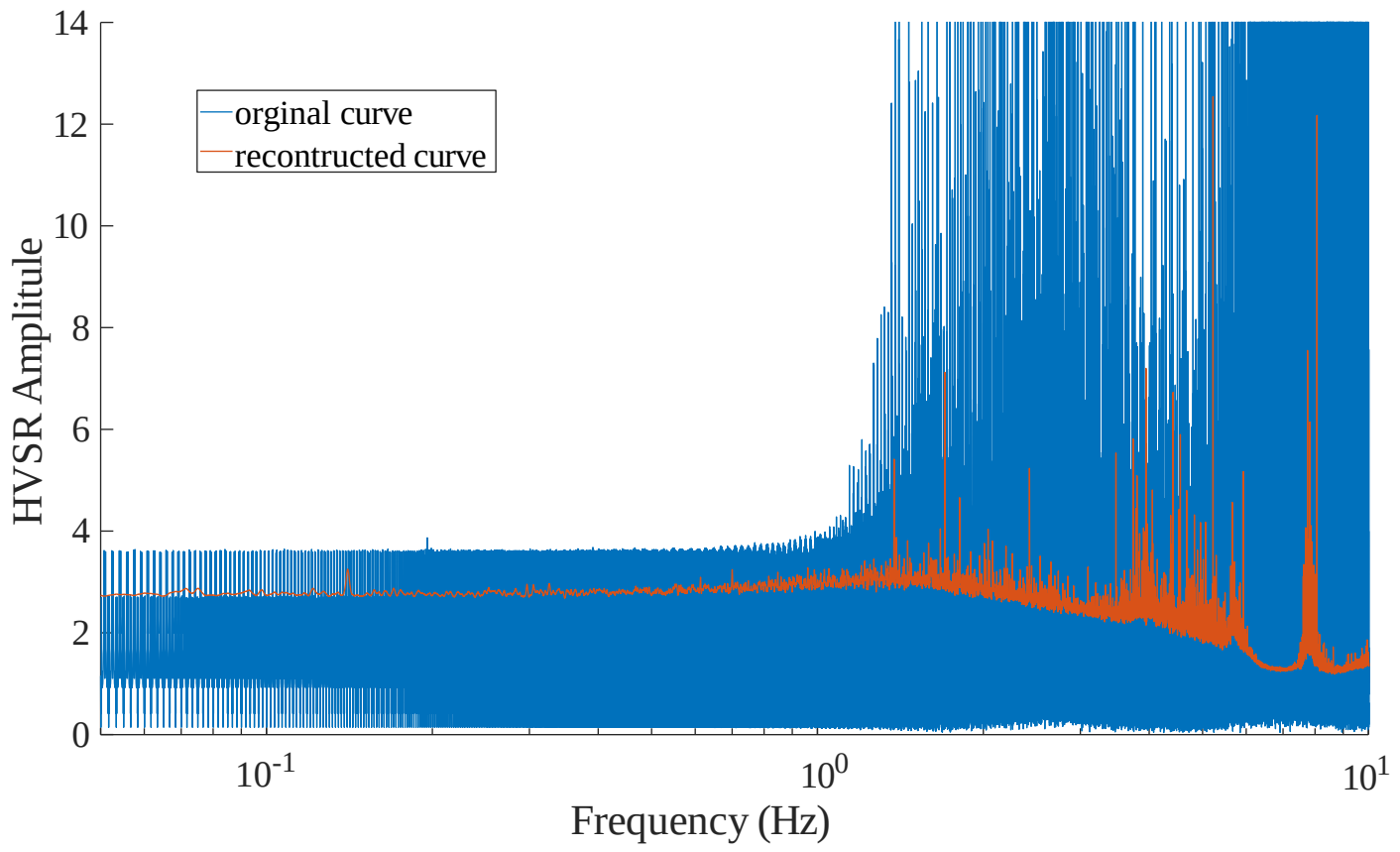

Figure 9: Orginal HVSR curve and recontructed HVSR curve

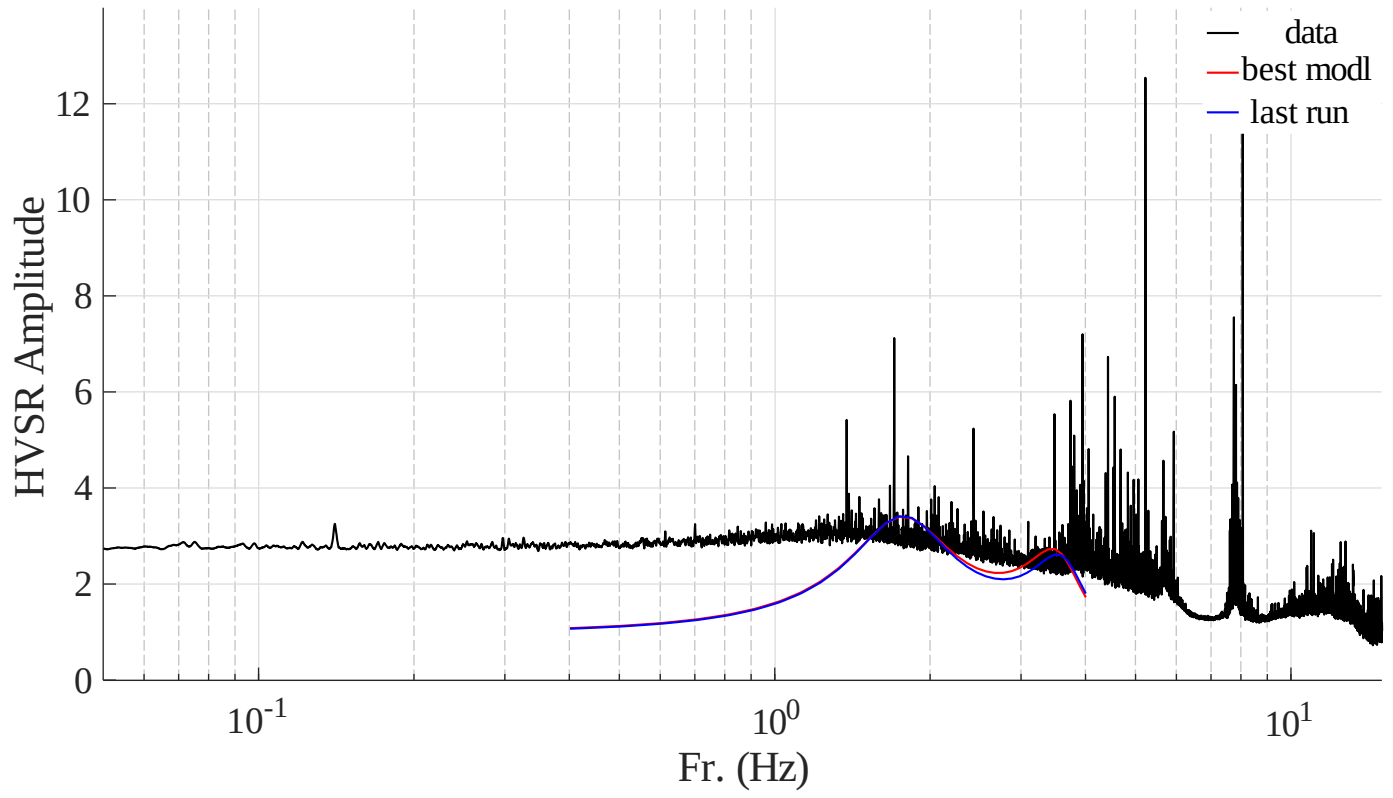

Figure 10: Inversion HVSR Curve

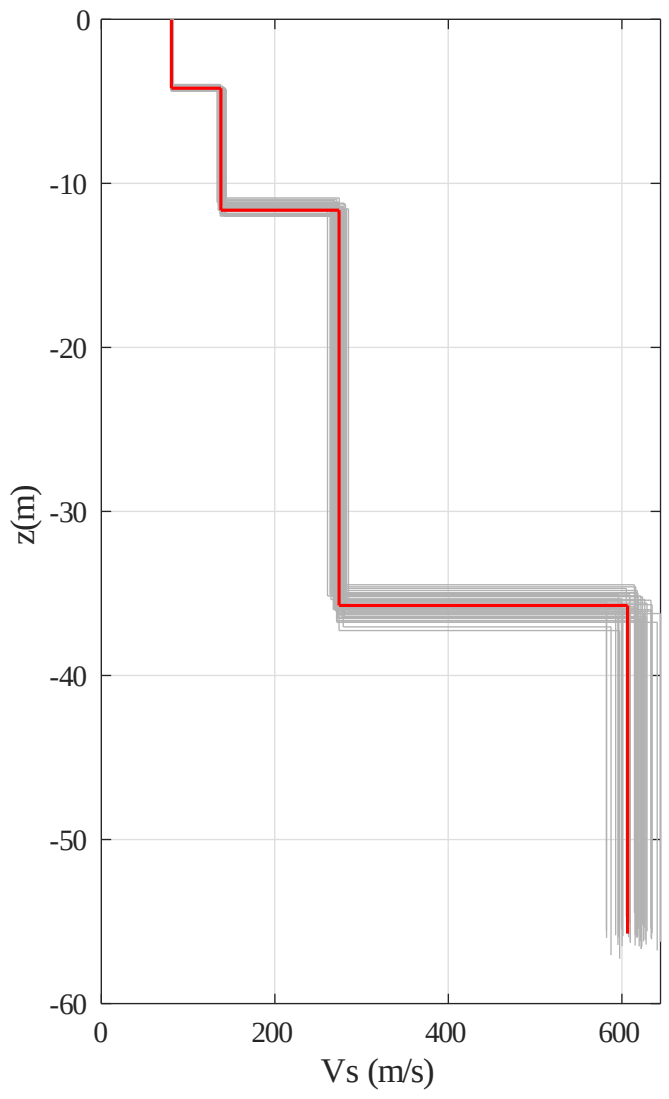

Figure 11: Inversion Vs

Then,correlation coefficients between $IMFs$ and orginal data are calculated, which are shown in Fig.6. The correlation coefficient values of $IMF$ $(1)$ and $IMF$ $(2)$ show that this example belongs to case 1 in the special cases. According to the filtering principle of case 1, we will filter out $IMF$ $(1)$. Then,the correlation coefficients with the first smoothed data are caculated, the results are shown in Fig.7. After that, the correlation coefficients with the second smoothed data are caculated, the results are shown in Fig.8. After the above processing, the effective signal is extracted.

The blue line in Fig.9 shows the spectral ratio curve of the strong interference data after preliminary processing, and it can be seen that the low-frequency effective signal is completely covered.And the red line in Fig.9 is the reconstructed curve.It shows that low-frequency effective signal is extracted. The micro-motion signal is reconstructed to extract information and inversion, and the results are shown in Fig.10 and Fig.11 .

The inversion results match well with the layer information revealed by the multi-electrode electrical measurement system (the first layer $(3m)$ : artificial soil and clay layer, the second layer ($3m \sim 18m$) : clay layer, the third layer ($18m \sim 25m$) : sand layer, and the fourth layer ($25m$ and below) :intensely weathered granite). Due to the uneven thickness of the sand layer revealed by the high-density electrical method, the location of this data may be located above the thick sand layer, and the weathering and fracture of the granite below lead to deeper stratification.

# 5 discussion

$n_{imf}$ is the numerical value that estimates the number of $IMFs$ generated after UPEMD.

When dealing with real data, it often occurs that multiple IMFs describe the decomposition of large amplitude interference trends of data. Therefore, the number of IMFs decomposed by UPEMD is artificially limited to less than the number of $n_{imf}$, and the residual signal (almost composed of large amplitude and long period interference events) is used as the residual output of UPEMD, which can reduce the amount of calculation while maintaining the quality of separated signals.

# 6 Conclusion

HVSR is a useful tool in urban exploration,However, HVSR measurements depends on random noise whitch carrying layers' information. In practice, the method suffers a lot from high energy industrial noise, which limts its range of application. In this paper, we first propose the neeed for strong interference data defination. Then,the HVSR curve reconstruction method is proposed for the denoising requirements of strong interference data in complex environments. The new method use UPEMD to decompose the orginal data,caculate the $IMFs'$ correlation ciefficient with smoothed data twice. Benefit from the above process, those remaining $IMFs$ are supposed to be valid signals. The application of the proposed method to the real data shows that the proposed method can extract data with strong interference and accurately invert the layers when the existing methods cannot.

# 7 Acknowledgments

The research work leading to this paper has been financially supported by The National Nature Science Foundation of China under grant no. 42074150 and Emergency Management Bureau Project of Futian District,Shenzhen:FTCG2023000209.

The authors are with the College of Geo-exploration Science and Technology,Jilin University,China,130026 (e-mail:A_box_of_hamsters@outlook.com; hanfx@jlu.edu.cn; chenyb2321@mails.jlu.edu.cn; wlj2784405913@outlook.com; huangmt2321@mails.jlu.edu.cn; panyjgryx@163.com).

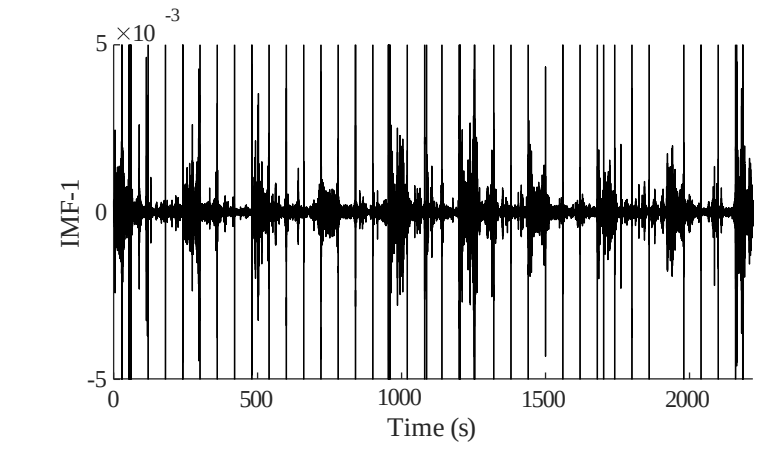

[1]

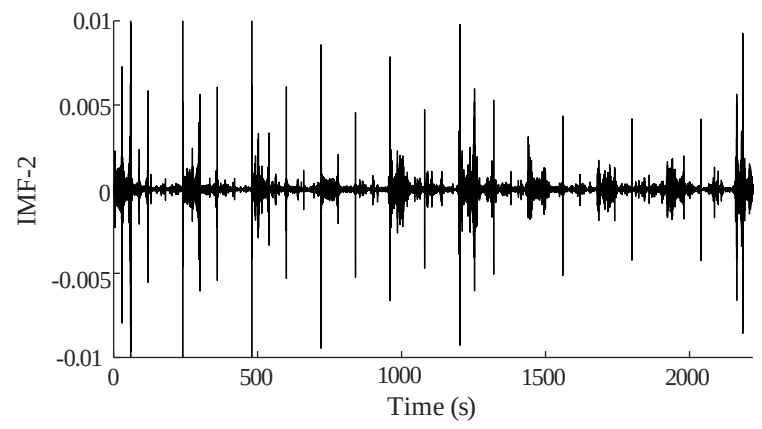

[2]

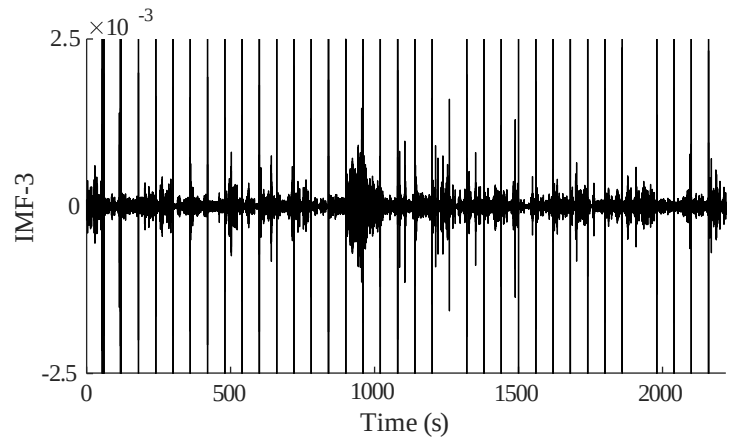

[3]

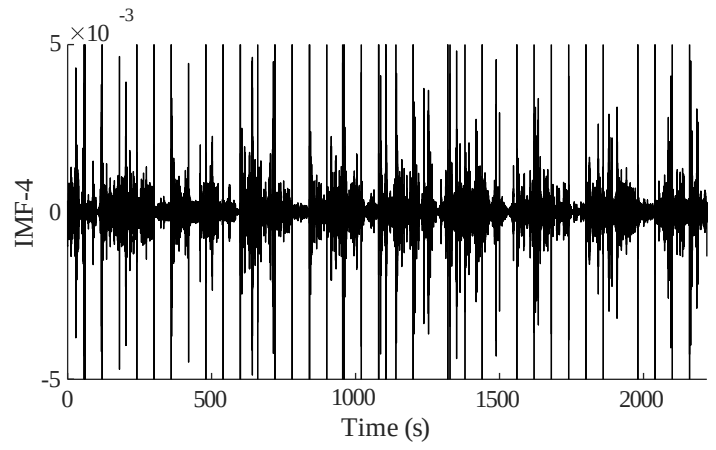

[4]

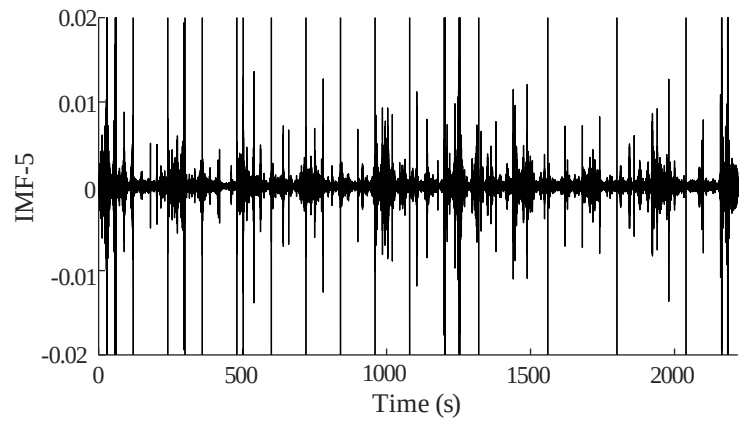

[5]

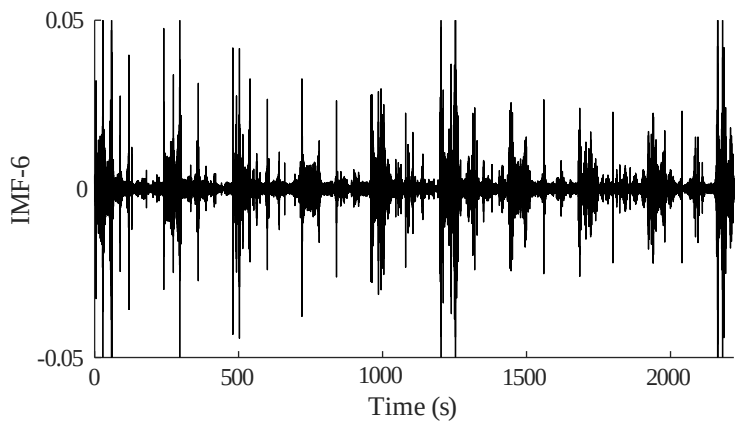

[6]

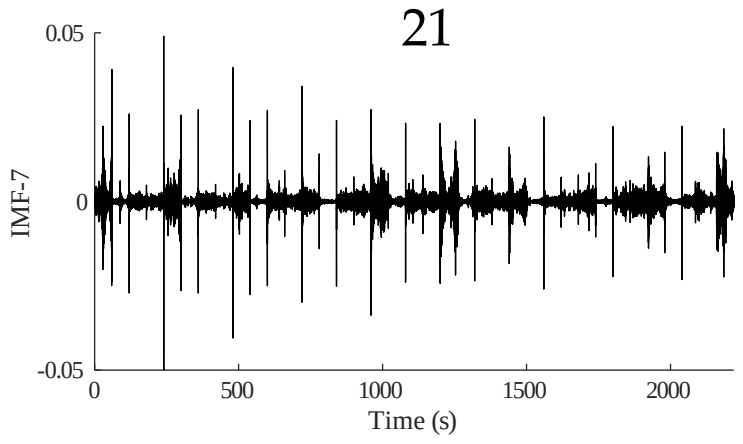

[7]

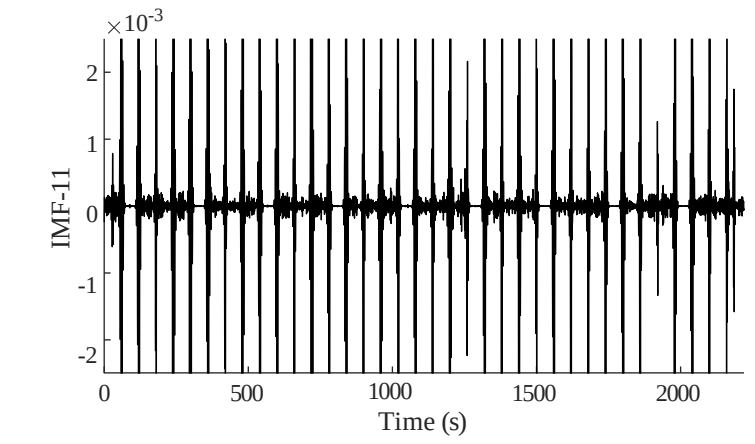

[11]

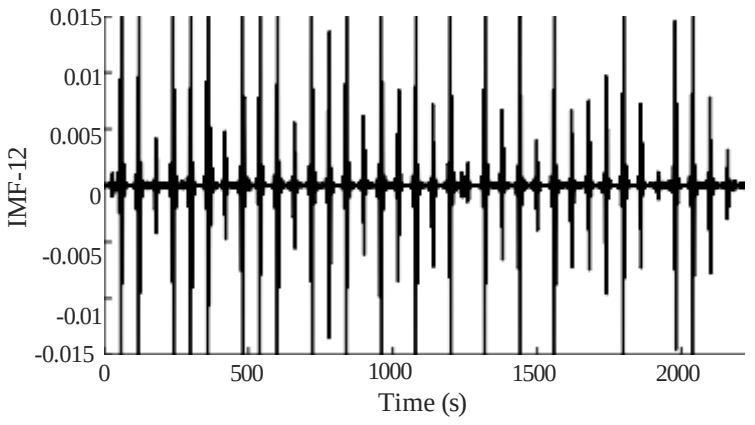

[12]

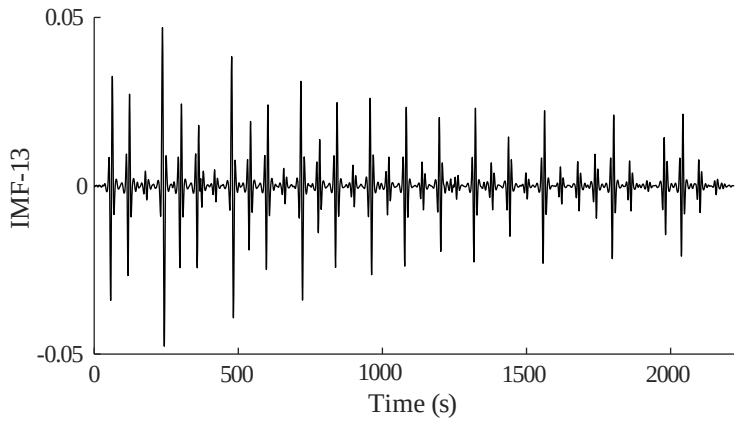

[13]

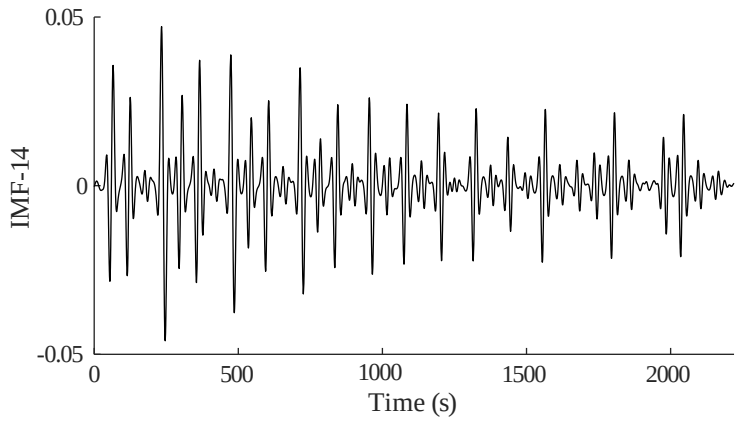

[14]

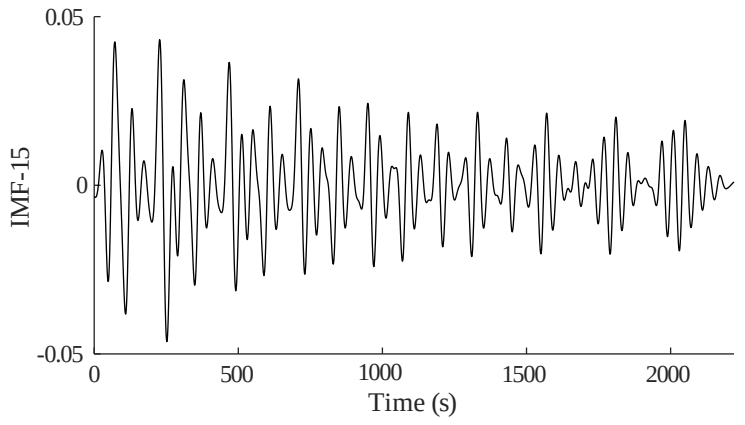

[15]

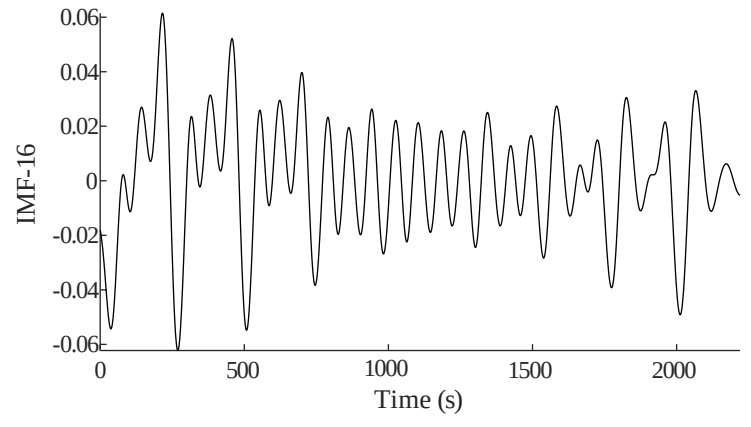

[16]

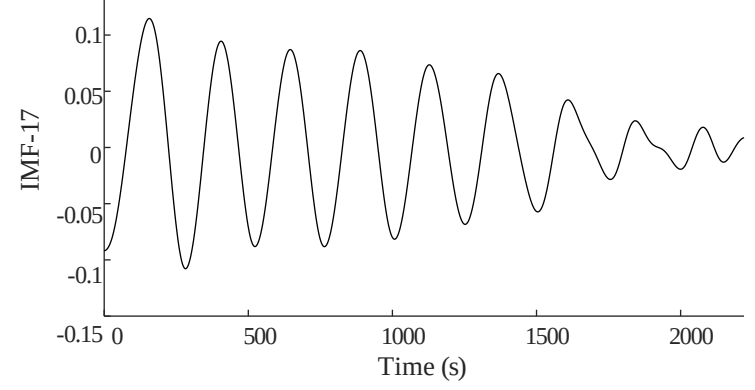

[17]